\documentclass[12pt]{article}

\usepackage{amsmath,amsfonts,amssymb}

\def\mH{\mathcal{H}}

\def\bx{\mathbf{x}}
\def\by{\mathbf{y}}

\newcommand{\tD}{\tilde{D}}
\newcommand{\mG}{\mathcal{G}}

\newcommand{\mL}{\mathcal{L}}

\def\pb #1{\left\{#1\right\}}

\begin{document}

\begin{titlepage}

\rightline{\footnotesize{CERN-PH-TH/2010-125}} \vspace{-0.2cm}

\begin{center}

\vskip 0.4 cm

\begin{center}
{\Large{ \bf Hamiltonian Analysis of the Higgs
Mechanism for Graviton}}
\end{center}

\vskip 1cm

{\large Josef Kluso\v{n}$^{}$\footnote{E-mail:
{\tt klu@physics.muni.cz}}
}

\vskip 0.8cm

{\it
Department of
Theoretical Physics and Astrophysics\\
Faculty of Science, Masaryk University\\
Kotl\'{a}\v{r}sk\'{a} 2, 611 37, Brno\\
Czech Republic\\
[10mm]
and
\vskip 10mm
Theory Division, Physics Department, CERN, \\
CH-1211 Geneva 23, Switzerland\\
}

\vskip 0.8cm

\end{center}

\begin{abstract}
In this paper we perform the canonical
description  of
the Higgs mechanism for gravity and provide
the Hamiltonian definition of the
massive gravities.
\end{abstract}

\bigskip

\end{titlepage}

\newpage

\section{Introduction}\label{first}
Construction of the massive gravity
is plagued with theoretical problems and
issues that have not been resolved from
the pioneered work of Fierz and Pauli
\cite{Fierz:1939ix}
\footnote{For extensive reviews of various
aspects of IR modification of gravities, see
\cite{Blas:2008uz,Bebronne:2009iy,Rubakov:2008nh,Vainshtein:2006ix}}.
One such a well known problem of massive gravity
is that there is no smooth massless limit in
the perturbation theory of massive gravity in
the sense that the massless limit in massive
gravity exists but does not agree with the
results derived from General Relativity (GR) that
describes massless gravitons. This pathological
behavior of massive gravity is known as
van Dam-Veltman-Zakharov discontinuity
\cite{vanDam:1970vg,Zakharov:1970cc}.

Even if  the construction of
 massive  gravity
is interesting theoretical challenge there is
another more stronger reason for their formulation.
In fact, generalizations
of GR with a small, non-zero,
 graviton mass leads to
 large scale (or
infrared) modifications of General Relativity.
Since GR has been directly tested from scales of
a fraction of millimeters up to Solar System scales
it is possible that the structure of the theory changes
on the large distances and hence
 infrared deviations from GR cannot be excluded.
Then the massive gravities could modify GR
at large cosmological scales and also can explain recent
accelerated expansion of the universe without assuming the
existence of mysterious dark matter and dark energy.

At present
it is not clear how to formulate  consistently
 a theory of massive gravity.  Since the
 Einstein's theory of gravity is well tested
 theory
 it seems to be  natural to add
 to the action of GR a term which will,
 in the linearized approximation,
give a mass to gravitons without
 modifying the kinetic terms coming from GR.

Very interesting formulation of the massive
theory of gravity is based on Brout-Englert-Higgs
mechanism for gravity. In more details,
we consider a diffeomorphism invariant action with usual
Einstein-Hilbert term together with the function
of the metric  coupled to $D$ scalar
fields for $D-$dimensional gravity
\cite{'tHooft:2007bf}. Then
gravitons acquire a mass
due to a mechanism which  may be thought of as  the Brout-Englert-Higgs
mechanism for gravity when
 the vacuum expectation value of each scalar field
  breaks one coordinate
reparameterization invariance
 \footnote{This interesting proposal of formulation
 of the massive gravity was also studied in
\cite{Kakushadze:2000zn,Kakushadze:2007dj,
Kakushadze:2007hf,Oda:2010gn,Chamseddine:2010ub,Oda:2010wn}.}.

Our goal is to develop the Hamiltonian
description of  the systems studied
in \cite{'tHooft:2007bf,Kakushadze:2000zn,Kakushadze:2007dj,
Kakushadze:2007hf,Oda:2010gn,Chamseddine:2010ub,Oda:2010wn}.
With the help of $(D-1)+1$ split
formalism for the space-time metric
 we find corresponding Hamiltonian
and show that it is linear combination of the Hamiltonian
and diffeomorphism constraints. The consistency of the theory
demands that these constraints are preserved during the
time evolution of the system. In order to
 check whether they are preserved or not
  we calculate the Poisson brackets of these constraints. However it turns out
that it is non-trivial task to calculate the Poisson
bracket of the Hamiltonian constraint for the scalar field
whose dynamics is governed by general action.
Despite of this fact we show that the Poisson brackets
of the scalar fields Hamiltonian constraints are
proportional to the diffeomorphism constraints.
In other words we   show that the Poisson bracket
of the scalar field Hamiltonian
constraints takes exactly the same form as the Poisson
bracket of General Relativity Hamiltonian constraints. Collecting
all these results  we obtain the Poisson
algebra of constraints is closed and hence these
constraints are consistent with the time-evolution
of the system. This result
is crucial for the possibility of the fixing
the gauge in the framework of the Hamiltonian
formalism. We fix the gauge
 by introducing $D$ gauge
fixed functions  that correspond to the
gauge fixing conditions introduced in
\cite{'tHooft:2007bf,Kakushadze:2000zn,Kakushadze:2007dj,
Kakushadze:2007hf,Oda:2010gn,Chamseddine:2010ub,Oda:2010wn}.
 Then the system of these gauge
fixing functions and original constraints form the set of
the second class constraints that can be explicitly solved.
 Further, since the
original Hamiltonian  is given as the linear
combination of constraints we find that in the process
of the gauge fixing  it strongly vanishes.
On the other hand the gauge fixing implies that the
reduced phase space is
 spanned by  the spatial components of
 the metric $h_{ij}$ and their
conjugate momenta $p^{ij}$. Then we  argue that the
the gauge fixing condition $t=\phi^0$ naturally
introduces the Hamiltonian on the reduced phase-space
equal to $-p_0$ where $p_0$ is the momentum conjugate
to $\phi^0$. Note that  $p_0$ is the function of reduced phase
space variables $h_{ij},p^{ij}$ as a result of the
solving of the Hamiltonian constraint. Say differently,
we claim that the definition of the massive
gravity is given by the reduced phase space variables
$h_{ij},p^{ij}$ together with the gauge fixed
Hamiltonian  $H_{fix}=-p_0(h_{ij},p^{ij})$,
and where all  
non-dynamical modes are absent.

Experiences from many areas of theoretical physics
teach us that in some cases the
 Lagrangian formulation of given theory is much more
 efficient then the Hamiltonian ones. For that
 reason we feel that it is useful to determine
 the Lagrangian formulation of the gauge fixed
 theory as well. It turns out that in order to
 find this Lagrangian
  it is convenient to  introduce new non-dynamical
modes so that the 
Legendre transformation from the Hamiltonian
to Lagrangian formulation can be easily 
performed.
Interestingly this Lagrangian can be written in such
a form that resembles
 the standard Einstein-Hilbert action together
with the potential term that explicitly breaks
the diffeomorphism invariance of given theory.

The structure of this paper is as follows. In the next
section (\ref{second})
 we perform the Hamiltonian analysis of the
 system introduced in
\cite{'tHooft:2007bf,Kakushadze:2000zn,Kakushadze:2007dj,
Kakushadze:2007hf,Oda:2010gn,Chamseddine:2010ub,Oda:2010wn}.
Then in section (\ref{third}) we calculate
of the algebra of constraints. Section (\ref{fourth})
is devoted to the study of the gauge fixed theory.
 In section (\ref{fifth})
we give two examples of  scalar potentials that
allow us to find explicit form of the gauge fixed Hamiltonian.
Finally in conclusion (\ref{sixth}) we outline our results and
suggest possible extension of this work.

\section{Hamiltonian Analysis of  General Relativity 
with Massless Scalars}\label{second}
In this section we develop the Hamiltonian
formalism for the following action
\begin{equation}\label{action}
S=\frac{1}{16\pi G}
\int d^D x \sqrt{-g}[R-L(H^{AB})] \ ,
\end{equation}
where $G$ is $D-$dimensional Newton's constant
and the induced internal metric is defined as
\begin{equation}
H^{AB}=g^{\mu\nu}\nabla_\mu \phi^A
\nabla_\nu \phi^B \ ,
\end{equation}
where $\phi^A $ are real $D$ scalar
fields with $A=0,\dots, D-1$ and
where the indices $A,B,\dots$ are raised and lowered
using the metric
$\eta_{AB}=\mathrm{diag}(-1,1,\dots,1)$. Finally,
$L$ is apriori a generic function of $H^{AB}$. The detailed
analysis  of properties of  function $L$ was performed in
\cite{Kakushadze:2007dj,Kakushadze:2007hf,Chamseddine:2010ub,Oda:2010wn}.
It is important to stress that the form of the potential
$L$ is not arbitrary. In fact, it was argued in
\cite{Kakushadze:2007dj,Kakushadze:2007hf,Chamseddine:2010ub,Oda:2010wn}
that this potential has to lead to the equations
of motion that possess following vacuum solution
\begin{equation}
\phi^A=x^\mu \delta_\mu^A \ , \quad
g_{\mu\nu}=\eta_{\mu\nu} \ .
\end{equation}
This requirement leads to the constraint on the
potential $L$
\begin{equation}
\frac{\delta L(H_*)}{\delta H^{AB}}=
\frac{1}{2}\eta_{AB} L(H_*) \ ,
\end{equation}
where $H_*^{AB}=\eta^{AB}$.

Our goal is to develop the Hamiltonian formalism for
system defined by the action
(\ref{action}). As usual in the study of the Hamiltonian
formalism for gravity we introduce
$(D-1)+1$ formalism.
 Explicitly, let us consider $D$ dimensional
manifold $\mathcal{M}$ with the
coordinates $x^\mu \ , \mu=0,\dots,D-1$
and where $x^\mu=(t,\bx) \ ,
\bx=(x^1,\dots,x^{D-1})$. We presume that
this space-time is endowed with the
metric $g_{\mu\nu}(x^\rho)$ with
signature $(-,+,\dots,+)$. Suppose that
$ \mathcal{M}$ can be foliated by a
family of space-like surfaces
$\Sigma_t$ defined by $t=x^0$. Let
$h_{ij}, i,j=1,\dots,D$ denotes the
metric on $\Sigma_t$ with inverse
$h^{ij}$ so that $h_{ij}h^{jk}=
\delta_i^k$. We further introduce the operator
$\nabla_i$ that is covariant derivative
defined with the metric $h_{ij}$.
 We  define  the lapse
function $N=1/\sqrt{-g^{00}}$ and
the shift function
$N^i=-g^{0i}/g^{00}$. In
terms of these variables we write the
components of the metric
$g_{\mu\nu}$ as
\begin{eqnarray}
g_{00}=-N^2+N_i h^{ij}N_j \ ,
\quad g_{0i}=N_i \ , \quad
g_{ij}=h_{ij} \ ,
\nonumber \\
g^{00}=-\frac{1}{N^2} \ , \quad
g^{0i}=\frac{N^i}{N^2} \ , \quad
g^{ij}=h^{ij}-\frac{N^i N^j}{N^2}
\ .
\nonumber \\
\end{eqnarray}
Then it is easy to see that
\begin{equation}
\sqrt{-g}=N\sqrt{\det h} \ .
\end{equation}
In the $(D-1)+1$ formalism
$H^{AB}$ takes the form
\begin{equation}
H^{AB}=-\nabla_n \phi^A\nabla_n
\phi^B+h^{ij}\partial_i \phi^A \partial_j \phi^B \ ,
\end{equation}
where
\begin{equation}
\nabla_n \phi^A=\frac{1}{N}(\partial_t\phi^A-N^i\partial_i
\phi^A) \ .
\end{equation}
Then from
 (\ref{action}) we easily  find
 the momenta conjugate to $\phi^A$
\begin{equation}
p_A=\frac{\sqrt{\det h}}{8\pi G}\frac{\delta L}{\delta H^{AB}}\nabla_n
\phi^B \ .
\end{equation}
Using this result we find following matrix
equation
\begin{eqnarray}\label{KD}
K_{AB}=D_{AC}(H^{AB})(V^{CD}-H^{CD})D_{DB}(H^{AB}) \ ,
\nonumber \\
\end{eqnarray}
where we introduced following matrices
\begin{eqnarray}
K_{AB}=\left(\frac{16\pi G}{2\sqrt{\det h}}\right)^2
p_A p_B \ , \quad  V^{AB}=h^{ij}
\partial_i \phi^A\partial_j\phi^B \ ,  \quad
D_{AB}(H^{AB})=\frac{\delta L}{\delta H_{AB}} \ .
\nonumber \\
\end{eqnarray}
For further purposes we also introduce
the matrix $\tD^{AB}$ inverse to
$D_{AB}$ so that
$D_{AB}\tD^{BC}=\delta_A^C$.

Now we presume that the  equation
(\ref{KD}) can be solved for $H_{AB}$. Let us
denote this solution as $J^{AB}=J^{AB}(K,V)$.
Then (\ref{KD}) implies
following  relation
\begin{equation}\label{VKD}
 (V^{AB}-J^{AB})=
\tD^{AC}(J)K_{CD}\tD^{DB}(J)
\end{equation}
that will be useful below.
Collecting all these results we find
the Hamiltonian of the scalar fields
in the form
\begin{equation}
H^\phi=\int d^{d} \bx (\partial_t \phi^A p_A-L)=
\int d^{d}\bx (N(\bx)\mH^\phi_T(\bx)+N^i(\bx)\mH^\phi_i(\bx)) \ ,
\end{equation}
where 
\begin{eqnarray}\label{Hamscal}
\mH^\phi_T&=&
\frac{1}{16\pi G}\left(2\sqrt{\det h}K_{AB}\tD^{BA}+V(J)
\right) \ , \nonumber \\
\mH_i^\phi&=&p_A\partial_i\phi^A \ , \nonumber \\
\end{eqnarray}
and where $d=D-1$. 

In the same way we should proceed with
the Hamiltonian analysis of the General Relativity
action. Since the procedure is well known
we immediately write the final result
\begin{equation}
H^{GR}=\int d^d\bx (N(\bx)\mH^{GR}_T(\bx)
+N^i(\bx)\mH^{GR}_i(\bx)) \ ,
\end{equation}
where
\begin{equation}\label{mHTGR}
\mH_T^{GR}=\frac{16\pi G}{\sqrt{\det h}}
\pi^{ij}\mG_{ijkl}\pi^{kl}-\frac{\sqrt{\det h}}{16\pi G}
R^{(D-1)} \ , \quad 
\mH_i=-2 h_{ik}\nabla_l \pi^{lk}
\ .
\end{equation}
Let us explain notation used here. $\pi^{ij}$
are momenta conjugate to $h_{ij}$.
The generalized metric $\mG^{ijkl}$  is
defined as
\begin{equation}
\mG_{ijkl}=\frac{1}{2}(h_{ik}h_{jl}
+h_{il}h_{jk})-h_{ij}h_{kl} \ .
\end{equation}
The inverse metric  $\mG^{ijkl}$ is equal to
\begin{equation}
\mG^{ijkl}=\frac{1}{2}(h^{ik}h^{jl}+
h^{il}h^{jk})-h^{ij}h^{kl} \ .
\end{equation}
Finally, $R^{(D-1)},\nabla_i $ are  Ricci curvature
and covariant derivative
 calculated
using the metric  $h_{ij}$.
Finally  due to the fact that the
action (\ref{action}) does not contain
time derivative of $N,N^i$ we find that
corresponding conjugate momenta $\pi_N,\pi_i$
are primary constraints of the theory
\begin{equation}
\pi_i\approx 0 \ , \quad
\pi_N\approx 0  \ .
\end{equation}
The condition of preservation of these
constraints implies the existence of the
secondary ones
\begin{equation}\label{constG}
\mH_T=\mH_T^{GR}+\mH_T^{\phi}\approx 0 \ , \quad 
\mH_i=\mH_i^{GR}+\mH_i^{\phi} \approx 0
\end{equation}
or their smeared form
\begin{equation}
H_T(M)=\int d^d\bx M(\bx)\mH_T(\bx) \ , \quad
H_S(M^i)=
\int d^d \bx M^i(\bx) \mH_i(\bx) \ ,
\end{equation}
where $M(\bx),M^i(\bx)$ are arbitrary
functions.
\section{Algebra of Hamiltonian Constraints}
\label{third}
As the next step we have to demonstrate the
stability of the constraints (\ref{constG}),
or equivalently, we have to show that these
constraints are preserved during the time
evolution of the system.
 In fact, the careful
analysis of these constraints was performed
in 70's in the geometrodynamics program
\cite{Kuchar:1974es,Isham:1984sb,Isham:1984rz,Hojman:1976vp}.
 It was argued there that for the consistency
of theory the Poisson brackets of constraints  should have
the form
\begin{eqnarray}\label{PBgeo}
\pb{H_S(N^i),H_S(M^i)}&=&H_S(N^i\partial_i M^j-M^i
\partial_i N^j) \ , \nonumber \\
\pb{H_S(N^i),H_T(M)}&=& H_T(N^i \partial_i M) \ ,
\nonumber \\
\pb{H_T(N),H_T(M)}&=&
H_S((M\partial_j N-N\partial_j  M)h^{ji}) \ .
\nonumber \\
\end{eqnarray}
The aim of this section is to show that
the Poisson brackets of
constraints (\ref{constG}) obey the algebra
(\ref{PBgeo}).
 In
fact,  it is well known that
the  General Relativity constraints
\begin{equation}
H_T^{GR}(N)=\int d^D\bx N(\bx)\mH^{GR}_T(\bx) \ ,
\quad H_S^{GR}(N^i)=\int d^D \bx N^i(\bx)
\mH_i^{GR}(\bx)
\end{equation}
obey the Poisson brackets relations
(\ref{PBgeo}). Further 
in case of  scalar field
diffeomorphism constraint we easily find
\begin{eqnarray}
\pb{H_S^\phi(N^i),H_S^\phi(M^i)}=
H^\phi_S(N^i\partial_i M^j-M^i\partial_i N^j) \ .
\nonumber \\
\end{eqnarray}
Then using the fact that  the mixed
Poisson brackets  $\pb{H^\phi_S(N^i),H^{GR}_S(M^i)}$
are trivially  zero we find that the
Poisson brackets of diffeomorphism constraints
obey the first equation in (\ref{PBgeo}).

As the next step we calculate the Poisson
bracket between  diffeomorphism
constraint $H_S(N^i)$ and $H_S^\phi(M)$.
Using following Poisson brackets
\begin{eqnarray}
\pb{H_S(N^i),\phi^A}&=&-N^i\partial_i\phi^A \ , \nonumber \\
\pb{H_S(N^i),p_A}&=&-\partial_i (N^ip_A) \ , \nonumber \\
\pb{H_S(N^i),h_{ij}}&=&-N^k\partial_k h_{ij}-
\partial_i N^k h_{kj}-h_{ik}\partial_j N^k \ , \nonumber \\
\pb{H_S(N^i),\sqrt{\det h}}&=&-N^k\partial_k\sqrt{\det h}-
\partial_k N^k\sqrt{\det h} \  \nonumber \\
\end{eqnarray}
we easily find
\begin{equation}
\pb{H_S(N^i),K_{AB}}=-N^k\partial_k K_{AB} \ ,
\quad \pb{H_S(N^i),V^{AB}}=-N^k \partial_k V^{AB} \ .
\end{equation}
Collecting all these results we obtain
\begin{equation}\label{pbHSHTphi}
\pb{H_S(N^i),\mH_T^\phi}=
-N^k\partial_k \mH_T^\phi-\partial_k N^k \mH_T^\phi \ .
\end{equation}
Finally using the fact that
$\pb{H_S^{GR}(N^i),H_T^{GR}(M)}=
H_T^{GR}(N^i\partial_i M)$ and the equation
(\ref{pbHSHTphi}) we find that
\begin{equation}
\pb{H_S(N^i),H_T(M)}=H_T(N^i \partial_i M) \ .
\end{equation}
As the last step we  calculate
 the Poisson brackets of the Hamiltonian
constraints. While it is well known that
the Poisson bracket of the GR
Hamiltonian constraints takes precisely the
form given in
(\ref{PBgeo})
it is far from obvious
that for  the system defined by the action
(\ref{action}) it holds as well. In case of
simple scalar action $\sim g^{\mu\nu}\partial_\mu \phi
\partial_\nu\phi$ the result is well known,
see for example
\cite{Kuchar:1974es,Isham:1984sb,Isham:1984rz,Hojman:1976vp}.
Further, it was shown very
elegantly in \cite{Kuchar:1974es}  that
the scalar field action in the form $L(-g^{\mu\nu}
\partial_\mu\phi\partial_\nu\phi)$ leads to
the Hamiltonian formulation where the constraints
obey the Poisson brackets (\ref{PBgeo}). On the other
hand the scalar field Hamiltonian
constraint (\ref{Hamscal}) is more general
then the form of the Hamiltonian constraints studied
in \cite{Kuchar:1974es} so that 
we perform the explicit calculation
of the Poisson brackets of the smeared
form of the Hamiltonian constraints
(\ref{Hamscal}) below.

Before we proceed to this calculation we discuss  the
calculation of the mixed Poisson brackets between
the GR Hamiltonian constraint  and scalar fields
Hamiltonian constraint.
The crucial point is that the
constraint $H^\phi_T$ depends on $g_{ij}$ only.
Then it is easy to see that
\begin{eqnarray}
& &\pb{H_T^{GR}(N),H_T^\phi(M)}+
\pb{H_T^\phi(N),H_T^{GR}(M)}=\nonumber \\
&=&\int d^D\bx d^D\by N(\bx)M(\by)
\left(\pb{\mH_T^{GR}(\bx),\mH_T^\phi(\by)}
+\pb{\mH_T^\phi(\bx),
\mH_T^{GR}(\by)}\right)=0 \nonumber \\
\end{eqnarray}
using the fact that $\pb{\mH^{\phi}_T(\bx),\mH^{GR}_T(\by)}
\sim \delta(\bx-\by)$.

\subsection{Poisson Brackets of Hamiltonian
Constraints for Scalar Field}
In this section we determine
the Poisson bracket between $H^\phi_T(N),H_T^\phi(M)$.
As a warm example we begin with the single scalar field
 action
\begin{equation}\label{SL}
S=\int d^D x \sqrt{-g}L(-I/2) \ ,
\end{equation}
where $I=g^{\mu\nu}
\partial_\mu\phi\partial_\nu\phi$.
In   $(D-1)+1$ formalism this action takes the form
\begin{equation}
S=\int d^D x N\sqrt{\det h}
L(\frac{1}{2}(\nabla_n^2\phi-h^{ij}\partial_i \phi
\partial_j\phi)) \ .
\end{equation}
Then it  is easy to see that  the momentum
 conjugate to $\phi$ is equal to
\begin{equation}\label{pinabla}
p=\sqrt{\det h}L'(I/2)\nabla_n\phi \
\end{equation}
and consequently
\begin{equation}\label{KI}
K=\left[\frac{p}{\sqrt{\det h}}\right]^2=L'^2(I/2)(I+V)  \ ,
\end{equation}
where $V=h^{ij}\partial_i \phi\partial_j \phi$
and where $L'(x)=\frac{dL(x)}{dx}$.
Let us presume that the equation (\ref{KI}) can
be solved for $I$ and  we denote this solution
as $I=J(K,V)$.
Then (\ref{pinabla}) implies
\begin{equation}
\nabla_n\phi=\frac{1}{\sqrt{h}L'(J/2)}
\pi
\end{equation}
and hence the Hamiltonian takes the form
\begin{equation}
H=\int d^d\bx( N(\bx)\mH_T(\bx)
+N^i(\bx)\mH_i(\bx) ) \ ,
\end{equation}
where
\begin{equation}
\mH_T= \frac{p^2}{\sqrt{\det h}L'(J/2)}
-\sqrt{\det h}L(J/2) \ .
\end{equation}
Our goal is to calculate
the Poisson bracket $\pb{H_T^\phi(N),H_T^\phi(M)}$.
To do this we start with the calculation
of the following Poisson bracket
\begin{eqnarray}\label{phiHT}
\pb{\phi,H_T}
&=&\int d^d\bx \pb{\phi,
N(\bx)\mH_T(\bx)}=
N\frac{2p}{\sqrt{\det h}}\frac{1}{L'^2(J/2)}-
\nonumber \\
&-&N\frac{p^3}{(\sqrt{\det h})^3}\frac{L''(J/2)}{L'^2(J/2)}
\frac{\delta J}{\delta K}
-N\frac{p}{\sqrt{\det h}}L'(J/2)
\frac{\delta J}{\delta K}  \ , \nonumber \\
\end{eqnarray}
where we used
\begin{eqnarray}
\pb{\phi(\bx),J(\by)}=
2\frac{p}{\det h}\frac{\delta J}{\delta K}(\bx)
\delta(\bx-\by) \ .
\nonumber \\
\end{eqnarray}
To proceed further we  note that $J$
obeys the equation
\begin{equation}\label{KJ}
K=L'^2(J/2)(J+V)  \ .
\end{equation}
When we difference this equation
with respect to $K$  and use
the fact that $J=J(K,V)$ we
find
\begin{eqnarray}
KL''(J/2)\frac{\delta J}{\delta K}=L'(J/2)
\left(1-L'^2(J/2)\frac{\delta J}{\delta K}\right) \ .
\nonumber \\
\end{eqnarray}
Inserting this result into  (\ref{phiHT})
and performing the appropriate manipulation
we find that
the Poisson bracket (\ref{phiHT})
takes the form
\begin{eqnarray}\label{phiHTphi}
\pb{\phi,H_T(N)}=
N\frac{\pi}{\sqrt{\det h}L'(J/2)}=N\nabla_n\phi
\nonumber \\
\end{eqnarray}
that agrees with the result
derived in \cite{Kuchar:1974es}.

As the next step  we determine the Poisson bracket
between $p$ and $H_T(N)$
\begin{eqnarray}\label{piHT}
& &\pb{p,H_T(N)}=
\int d^d\bx N(\bx)\times \nonumber \\
&\times & \left(-\frac{p^2}{2\sqrt{\det h}L'^2(J/2)}
L''(J/2)\frac{dJ}{dV}(\bx)\pb{p,V(\bx)}
-\frac{1}{2}\sqrt{\det h}L'(J/2)\frac{\delta J}{dV}(\bx)
\pb{p,V(\bx)}\right) \ .
\nonumber \\
\end{eqnarray}
With the help of the  relation (\ref{KJ})
we find
\begin{equation}
KL''(J/2)\frac{dJ}{dV}=-L'^3(J/2)
\left(1+\frac{dJ}{dV}\right) \ , \quad
\frac{dJ}{dV}=-L'\frac{dJ}{dK}
\ .
\end{equation}
Inserting these expressions into (\ref{piHT})
we find
\begin{eqnarray}\label{piH1}
\pb{p,H_T(N)}=
\frac{1}{2}\int d^d\bx N
\sqrt{\det h}L'(\bx)\pb{p,V(\bx)}=
\partial_i[N\sqrt{\det h}h^{ij}L'
\partial_j\phi] \ ,
\nonumber \\
\end{eqnarray}
where we also used
\begin{equation}
\pb{p(\bx),V(\by)}=
-2h^{ij}(\by)\partial_{y^j}\phi(\by)
\partial_{y^i}\delta(\bx-\by) \ .
\end{equation}
Now we are ready to determine
the Poisson bracket $\pb{H_T(N),
H_T(M)}$. To do this we
consider following expression
\begin{equation}\label{difpb}
\pb{\pb{\phi,H_T(N)},
H_T(M)}-\pb{\pb{\phi,H_T(M)},
H_T(M)} \ .
\end{equation}
In the first step we use  (\ref{phiHTphi}) and
 we find
\begin{equation}\label{pbhelp}
\pb{N\frac{\pi}{\sqrt{\det h}L'(J/2)},H_T(M)}
-\pb{M\frac{\pi}{\sqrt{\det h}L'(J/2)},H_T(N)} \ .
\end{equation}
Clearly all terms in (\ref{pbhelp})
that are not proportional to the partial derivatives
of $N,M$ cancel. To proceed further in
the calculation of  (\ref{pbhelp})
we  have to calculate
 following Poisson bracket
\begin{eqnarray}
\pb{\frac{1}{L'},H_T(N)}
&=&-2N\frac{L''(J/2)}{L'^2(J/2)}
\frac{dJ}{dK}\times \nonumber \\
&\times &\left
(-L'(J/2)h^{ij}\partial_i\phi
\partial_j[\frac{p}{\sqrt{\det h}L'(j/2)}]
+ \frac{p}{\det h}\partial_i[\sqrt{\det h}h^{ij}
\partial_j\phi]\right) \ .
\nonumber \\
\end{eqnarray}
Then it is easy to see
that  $\frac{\pi}{\sqrt{\det h}}\left(
M\pb{\frac{1}{L'},H_T(N)}-N
\pb{\frac{1}{L'},H_T(M)}\right)=0$.
Finally with  the help of
(\ref{piH1}) we find that
(\ref{difpb}) is equal to
\begin{eqnarray}
\pb{\pb{\phi,H_T(N)},H_T(M)}
-\pb{\pb{\phi,H_T(M)},H_T(N)}
=(N\partial_i M-M\partial_i N)h^{ij}\partial_j \phi \ .
 \nonumber \\
\end{eqnarray}
With the help of the Jacobi identity
we can rewrite this expression 
into the form
\begin{eqnarray}
\pb{\phi,\pb{H_T(N),
H_T(M)}}&=&(N\partial_i M-M\partial_i N)h^{ij}\partial_j \phi
=\nonumber \\
&=&\pb{\phi,H_S(N\partial_i M-M\partial_i h^{ij})+f(\phi)} \
\nonumber \\
\end{eqnarray}
for arbitrary  function $f(\phi)$. On the other
hand  performing the same step with $\pi$ we find
that $f$ does not depend on $\phi$ as well and
hence could be taken to vanish at least   at classical level.
In summary, we proved that the Poisson bracket
of the Hamiltonian constraints of the
scalar field with general action
(\ref{SL})   takes the
desired form
\begin{equation}
\pb{H_T(N),
H_T(M)}=
H_S((N\partial_i M-M\partial_i N) h^{ij}) \ .
\end{equation}
Now we are ready to proceed to
the calculation of the Poisson bracket
of the  Hamiltonian
constraint when the action for the scalar
fields is given in
 (\ref{action}).
We again start with the calculation
of the Poisson bracket between $\phi^X$ and
$H_T^\phi$
\begin{eqnarray}\label{phiXHh}
\pb{\phi^X,H_T^\phi}&=&
N\sqrt{\det h}\left(2\pb{\phi^X,K_{AB}}\tD^{BA}-\right.\nonumber \\
&-& \left. 2K_{AB}\tD^{BC}\frac{\delta D_{CD}}{\delta K_{MN}}
\tD^{DA}\pb{\phi^X,K_{MN}}
+D_{AB}\frac{\delta J^{BA}}{\delta K_{MN}}
\pb{\phi^X,K_{MN}}\right) \ .
\nonumber \\
\end{eqnarray}
To proceed we note that the equation
(\ref{VKD}) implies following
relation
\begin{equation}\label{KDC}
K_{AB}=D_{AC}(V^{CD}-J^{CD})D_{DB} \  ,
\end{equation}
where  $D_{AB}$ depends on $J$.
Then we calculate the Poisson bracket
between $\phi^X$ and the relation given
above, multiply the result with
 with $\tD^{BA}$ and finally take
 the traces over capital indices. As
 a result we find
\begin{eqnarray}
\tD^{BA}\pb{\phi^X,K_{AB}}=
2\tD^{BA}\frac{\delta D_{AC}}{\delta K_{MN}}
\pb{\phi^X,K_{MN}}\tD^{CD}K_{DB}
+\frac{\delta J^{BD}}{\delta K_{MN}}D_{DB}
\pb{\phi^X,K_{MN}} \ .
\nonumber \\
\end{eqnarray}
Using this result in (\ref{phiXHh})
we find
\begin{eqnarray}
\pb{\phi^X,H_T^\phi(N)}=
N\frac{1}{2\sqrt{\det h}}\tD^{XB}p_B=N\nabla_n\phi^X \ .
\nonumber \\
\end{eqnarray}
In the similar  way we proceed with the calculation
of  the Poisson
bracket between $p_X$ and $H_T^\phi$
\begin{eqnarray}
\pb{p_X,H^\phi_T(N)}=
2\partial_i[\sqrt{\det h}
h^{ij}N D_{XA}\partial_j \phi^A] \ ,
\nonumber \\
\end{eqnarray}
where we used
the Poisson bracket between $p_X$ and
(\ref{KDC}) that implies
\begin{equation}
2K_{PB}\tD^{BA}\frac{\delta D_{AC}}{\delta J^{MN}}
\frac{\delta J^{MN}}{\delta V^{PQ}}
\pb{p_X,V^{PQ}}=
\frac{\delta J^{CD}}{\delta V^{PQ}}D_{DC}\pb{p_X,V^{PQ}}
-\pb{p_X,V^{CD}}D_{DC} \ .
\end{equation}
As in case of single scalar field we
consider following expression
\begin{equation}
\pb{\pb{\phi^A,H^\phi_T(N)}H^\phi_M(M)}
-\pb{\pb{\phi^A,H^\phi_T(M)}H^\phi_T(N)} \ ,
\end{equation}
where the first double Poisson bracket
is equal to
\begin{eqnarray}\label{pbphiAHNM}
& &\pb{\pb{\phi^A,H^\phi_T(N)}H^\phi_M(M)}=
\frac{N}{\sqrt{\det h}}\tD^{AB}
\partial_i[M\sqrt{\det h}h^{ij}D_{BC}\partial_j\phi^C]-
\nonumber \\
&-&\frac{N}{\sqrt{\det h}}
\tD^{AC}\left(
\frac{\delta D_{CD}}{\delta J^{MN}}
\frac{\delta J^{MN}}{\delta K_{PQ}}
\pb{K_{PQ},H_T^\phi(M)}
+\frac{\delta D_{CD}}{\delta J^{MN}}
\frac{\delta J^{MN}}{\delta V^{PQ}}
\pb{V^{PQ},H_T^\phi(M)} \right)\tD^{DB}p_B \ .
\nonumber \\
\end{eqnarray}
To proceed further we have to 
determine  the
 Poisson bracket between $D_{AB}$
and $H_T^\phi(N)$.
Note that when 
 we  compare the variation of
(\ref{KDC}) with respect to $K_{MN}$
with the variation of (\ref{KDC}) with
respect to $V_{MN}$ we find following relation
\begin{equation}
\frac{\delta J^{MN}}{\delta V^{PQ}}=-
\frac{\delta J^{MN}}{\delta K_{RS}}
D_{RP}D_{SQ} \ .
\end{equation}
Then using this relation in (\ref{pbphiAHNM}) and
after some manipulation  we
find
\begin{equation}
\pb{\pb{\phi^A,H^\phi_T(N)}H^\phi_T(M)}-
\pb{\pb{\phi^A,H^\phi_T(M)}H^\phi_T(N)}
=(N\partial_i M-M\partial_i N)h^{ij}
\partial_j\phi^A \ .
\end{equation}
Performing the same analysis
with conjugate momenta $p_A$
we find  the desired result
\begin{equation}
\pb{H^\phi_T(N),H^\phi_T(M)}=
H^\phi_S((N\partial_i M
-M\partial_i N)h^{ij}) \ .
\end{equation}
The upshot of this long analysis is
the proof that the Poisson brackets
of  Hamiltonian
constraints $H_T(N)=H_T^{GR}(N)+
H_T^\phi(N)$ take exactly the same
form as in (\ref{PBgeo}). Using
this fact we immediately obtain
that the Hamiltonian and
diffeomorphism constraints are preserved
during the time evolution of the
system. This is very important
result since only after determining
the complete constraint structure
of given theory it is possible to
perform the Hamiltonian gauge fixing.
\section{Fixing Gauge}\label{fourth}
In this section we fix the gauge
freedom that in the Hamiltonian
treatment are expressed by an existence
of $D$ first class constraints
$\mH_T(\bx)\approx 0 , \mH_i(\bx)\approx 0$.
This procedure is
an analogue of the Higgs mechanism
used in the construction of the
massive gravity
\cite{'tHooft:2007bf,Kakushadze:2007dj,
Kakushadze:2007hf,Oda:2010gn,Chamseddine:2010ub}.
In these models the vacuum expectation
values of the scalar fields  coincide 
with $D$ space-time coordinates. The
result of this fixing is the complete
breaking of  the space-time diffeomorphism and
emergence of the mass term for the 
graviton in the action when we study the small 
fluctuations of gravity above the flat
space-time.

The standard way how to fix
the gauge freedom in Hamiltonian framework
is to introduce
 $D$ gauge
fixed functions that can
be interpreted as additional
constraints imposed on the system
and that have non-zero Poisson brackets
with the original first class constraints.
As a result the extended system of original
 constraints together with
gauge fixed functions form the collection
of the second class constraints
with no gauge freedom left
(For review of this formalism, see
\cite{Govaerts:2002fq,Govaerts:1991gd,Henneaux:1992ig}.).

With analogy with the fixing the gauge given in
\cite{'tHooft:2007bf,Kakushadze:2007dj,
Kakushadze:2007hf,Oda:2010gn,Chamseddine:2010ub}
we introduce  following $D$ gauge fixing
functions
\begin{equation}\label{GaugefixingF}
\mG^A(\bx)=\phi^A(\bx)-x^A \ .
\end{equation}
Clearly
\begin{equation}\label{pbGHi}
\pb{\mG^A(\bx),\mH_i(\by)}=
\delta_i^A\delta(\bx-\by)
\end{equation}
and
\begin{equation}\label{pbGH0}
\pb{\mG^A(\bx),\mH_T(\by)}=
\frac{(16\pi G)^2}{2\det h}\frac{\delta \mH_T^\phi(\bx)}{\delta
K_{AN}(\bx)}p_N(\bx)\delta(\bx-\by) \ .
\end{equation}
Due to the fact that the Poisson brackets
(\ref{pbGHi}) and (\ref{pbGH0}) are non-zero
on the constraint surface we see
that the collection of constraints
$(\mG^A,\mH_T,\mH_i)$ is the system of the
second class constraints. Alternatively,
the requirement of the time preservation of
the constraints $\mG^A(\bx)\approx 0$ during the
time evolution of the system implies
following consistency equation
\begin{eqnarray}
\frac{d\mG^i(\bx)}{dt}&=&
\pb{\mG^i(\bx),H}=\nonumber \\
&=&N^i(\bx)+N(\bx)\frac{(16\pi G)^2}{2\det h}
\frac{\delta H^\phi_T(\bx)}{\delta K_{iN}(\bx)}
p_P(\bx)=0 \ ,  \nonumber \\
\frac{d\mG^0(\bx)}{dt}&=&
\partial_t \mG^0(\bx)+\pb{\mG^0(\bx),H}=\nonumber \\
&=&
-1+N(\bx)\frac{(16\pi G)^2 }{2\det h}
\frac{\delta H^\phi_T(\bx)}{\delta K_{0N}(\by)}
\tD^{NP}p_P(\bx)=0 \ ,
\nonumber \\
\end{eqnarray}
where we used  (\ref{pbGHi}) and (\ref{pbGH0}). We see
that these equations determine
 $N$ and $N^i$ as functions of the canonical
 variables $h_{ij},p^{ij}$ with no gauge freedom left.

 The fact that $\mG^A,\mH_T,\mH_i$ are the second
 class constrains implies that they vanish strongly
 and can be explicit solved for $p_A$. As a result
 the reduced phase space is spanned by $h_{ij}$ and
 $p^{ij}$. Further,
 since the Hamiltonian of the original system
 was given as a linear combination of the constraints
 $\mH_T,\mH_i$ we now see that it vanishes strongly.

  On the other
hand let us write the original action (\ref{action})
in the form
\begin{equation}\label{action2}
S=\int dt d^d\bx (\partial_t h_{ij}\pi^{ji}
+\partial_t \phi^A p_A-H)=
\int dt (\int d^d\bx (\partial_t h_{ij}
\pi^{ji}+p_0(p^{ij},g_{ji}))) \ ,
\end{equation}
where we used the fact that $H=0$ and imposed
the gauge fixing functions (\ref{GaugefixingF}).
We see from (\ref{action2}) that it
is natural to  interpret $-p_0$ as the
Hamiltonian density of the reduced theory
\begin{equation}\label{Hredtheo}
\mH_{fix}=
-p_0(h_{ij}(\bx),\mH_T^{GR}(\bx),\mH^{GR}_i(\bx))\equiv
-\frac{16\pi G}{2\sqrt{\det h}}\tilde{p}_0 \ ,
\end{equation}
where we also used the fact that from
the  constraints
$\mH_i=0$ we can express $p_i$ as
\begin{equation}
p_i(\bx)=-\mH^{GR}_i(\bx) \ .
\end{equation}
Finally note  that $p_0$ can be derived
from the Hamiltonian constraint
$\mH_T=\mH_T^{GR}+\mH_T^\phi(\bx)=0$
at least in principle.

In summary, we found the Hamiltonian formulation
of massive gravity where the physical degrees
of freedom are $h_{ij},p^{ij}$ and where
the Hamiltonian is given in (\ref{Hredtheo}).
Note also that the  explicit form of this Hamiltonian
depends on the form of  the function $L(H^{AB})$.
We give simple examples of two solvable potentials  in
the next section. Generally however it is very
difficult to find explicit form of the gauge
fixed Hamiltonian due to the complicated structure
of the function $L(H^{AB})$.

Despite of this fact we now show
that  it is possible to find
the Lagrangian density for given gauge
fixed theory.
 To to this we introduce
four modes $A,B,C_i, D^i$ and corresponding
conjugate momenta $(p_A,p_B,p^i,p_i)$
with non-zero Poisson brackets
\begin{eqnarray}
\pb{A(\bx),p_A(\by)}&=&\delta(\bx-\by) \ ,
\quad \pb{B(\bx),p_B(\by)}=
\delta(\bx-\by) \ , \nonumber \\
\pb{C_i(\bx),p^j(\by)}&=&\delta_i^j
\delta(\bx-\by) \ , \quad
\pb{D^i(\bx),p_j(\by)}=\delta^i_j\delta(\bx-\by) \  .
\nonumber \\
\end{eqnarray}
With the help of these additional
modes we rewrite the Hamiltonian for
gauge fixed theory as
\begin{equation}\label{Hfixext}
\mH_{fix}=
-
p_0(h_{ij},A,C_i)+
B(\mH_T^{GR}-A)+D^i (\mH^{GR}_i-C_i)+
v_A p_A+v_B p_B+v^i p_i+v_i p^i \ ,
\end{equation}
where the Lagrange multipliers $
v_A,v_B,v^i,v_i$ ensure that $p_A,p_B,
p_i$ and $p^i$ are primary constraints
of the theory
\begin{equation}
p_A\approx 0 \ , \quad  p_B\approx 0 \ ,
\quad p_i\approx 0 \ , \quad  p^i\approx 0 \ .
\end{equation}
Then the fact that these constraints
have to be preserved during the time
evolution of the system implies the
secondary constraints
\begin{eqnarray}
\partial_t p_A&=&\pb{p_A,H_{fix}}=\frac{\delta p_0}{
\delta A}-B\equiv \Phi_A\approx 0 \ , \nonumber \\
\partial_t p_B&=&\pb{p_A,H_{fix}}=-A+\mH^{GR}_T\equiv \Phi_B\approx 0 \ , 
\nonumber \\
\partial_t p_i&=&\pb{p_i,H_{fix}}=-C_i+
\mH^{GR}_i\equiv \Phi_i\approx 0 \ ,
\nonumber \\
\partial_t p^i&=&\pb{p^i,H_{fix}}=
\frac{\delta p_0}{
\delta C_i}-D^i\equiv \Phi^i \approx 0 \ .
\nonumber \\
\end{eqnarray}
 It can be shown that the collections of the
 constraints $(p_A,p_B,p_i,p^i,\Phi_A,\Phi_B,\Phi^i,
 \Phi_i)$ are the second class constraints.
 Solution of these constraints reduces (\ref{Hfixext})
 into the
 original form of the gauge fixed Hamiltonian
 (\ref{Hredtheo}).

The main advantage of the extended form of the
Hamiltonian density
 (\ref{Hfixext}) is that it allows us to find  corresponding
 Lagrangian in relatively straightforward way.
 In fact, from (\ref{Hfixext}) we
 easily obtain the
 time derivatives of the canonical variables
 $h_{ij},A,B,D^i,C_i$
\begin{eqnarray}\label{parthij}
\partial_t h_{ij}&=&\pb{h_{ij},H_{fix}}=
 B\frac{32\pi G}{\sqrt{\det h}}\mG_{ijkl}\pi^{kl}+2\nabla_i D_j \ ,
\nonumber \\
\partial_t A&=&\pb{A,H_{fix}}=v_A \ ,
\quad \partial_t B=\pb{B,H_{fix}}=
v_B \ , \nonumber \\
\partial_t D^i&=&\pb{D^i,H_{fix}}=v^i \ , \quad
\partial_t C_i=\pb{C_i,H_{fix}}=v_i \ .   \nonumber \\
\end{eqnarray}
It turns out that it is useful to   introduce following object
\begin{equation}
\hat{K}_{ij}=\frac{1}{2B}(\partial_t h_{ij}
-\nabla_i D_j-\nabla_j D_i)
\end{equation}
that due to the first equation in
(\ref{parthij}) is related to $\pi^{ij}$
as
\begin{equation}
\hat{K}_{ij}=\frac{16\pi G}{\sqrt{\det h}}
\mG_{ijkl}\pi^{kl} \ .
\end{equation}
Then it is easy to find corresponding Lagrangian
\begin{eqnarray}\label{Lfixfinal}
L_{fix}
&=&\int d^d\bx \left(\frac{\sqrt{\det h}}{16\pi G}
B(
\hat{K}_{ij}\mG^{ijkl}\hat{K}_{kl}
+R^{(D-1)})
+\frac{\sqrt{\det h}}{8\pi G}
\tilde{p}_0(h_{ij},A,C_i)
+BA+D^iC_i\right)=
\nonumber \\
&=&\int d^d\bx \left(\frac{1}{16\pi G}
\sqrt{-\hat{g}}\hat{R}
+\frac{\sqrt{\det h}}{8\pi G}
\tilde{p}_0(h_{ij},A,C_i)
+BA+D^iC_i\right) \ ,
\nonumber \\
\end{eqnarray}
where $\hat{g}$ is $D-$dimensional
metric with components
\begin{equation}
\hat{g}_{00}=-B^2+D_ih^{ij}D_j \ , \quad \hat{g}_{0i}=D_i \ ,
\quad \hat{g}_{ij}=h_{ij} \ .
\end{equation}
and where $\hat{R}$ is $D+1$ Ricci scalar built from
this metric.  We see that the last form
of the Lagrangian (\ref{Lfixfinal})
can be interpreted as the sum of the General Relativity
action with additional potential terms that
breaks the full diffeomorphism invariance of the
theory. It is important to stress that this
potential term depends on the auxiliary fields
$A$ and $C_i$. In principle these terms could
be integrated out however we expect that the
resulting Lagrangian would be very complicated.
\section{ Examples of
Potentials $L(H^{AB})$}\label{fifth}
In this section we give two solvable examples
of the scalar function $L(H^{AB})$ that allow
to find the explicit form of Lagrangian
for massive gravity.

In the first case
we follow  \cite{'tHooft:2007bf}
and consider  function $L(H^{AB})$ in the form
\begin{equation}\label{Lhooft}
L=\Lambda+H_{AB}\eta^{BA} \ . 
\end{equation}
Then it is easy to see that
\begin{equation}
\frac{\delta L}{\delta H_{AB}}=\eta_{BA} \  
\end{equation}
and 
\begin{equation}
J^{AB}\eta_{BA}=V^{AB}\eta_{BA}-K^{AB}\eta_{BA} \ . 
\end{equation}
Using these results it is straightforward 
 exercise to find scalar field Hamiltonian
 density
\begin{equation}
\mH_T^\phi=\frac{\sqrt{\det h}}
{16\pi G}(K_{AB}\eta^{BA}+V^{AB}\eta_{BA}+\Lambda) \ .
\end{equation}
Then following the general procedure outlined 
in previous section we find
\begin{equation}\label{mHfixsim}
\mH_{fix}=\frac{\sqrt{\det h}}{8\pi G}
\sqrt{\frac{16\pi G}{\sqrt{\det h}}\mH_T^{GR}+
\left(
\frac{8\pi G}{\sqrt{\det h}}\right)^2\mH^{GR}_i\delta^{ij}\mH^{GR}_j+
h^{ij}\delta_{ji}+\Lambda} \ .
\end{equation}
Finally we find the
Lagrangian density of
the gauge fixed theory in the form
\begin{eqnarray}\label{mLfix2}
\mL_{fix}&=&\frac{\sqrt{\det h}}{16 \pi G}
B(\hat{K}_{ij}\mG^{ijkl}\hat{K}_{kl}+
R^{(D-1)}+\Lambda)+D_i C^i+AB
-\nonumber \\
&-&\frac{\sqrt{\det h}}{8\pi G}
\sqrt{A+
\left(
\frac{(8\pi G}{\sqrt{\det h}}\right)^2C_i\delta^{ij}C_j+
h^{ij}\delta_{ji}}
\ . \nonumber \\
\end{eqnarray}
We would like to stress that
we can integrate out auxiliary fields
$A,B,C^i,D_i$ from the Lagrangian
(\ref{mLfix2}) and then to derive
the Lagrangian density for dynamical
modes $h_{ij}$ only. However the resulting
Lagrangian would be very complicated and
hence we prefer to work with the extended
Lagrangian (\ref{mLfix2}).

As the second example of exactly solvable
theory we consider the Lagrangian function
$L(H^{AB})$ in the form
\begin{equation}\label{LH2ex}
L(H_{AB})=\Lambda+\sqrt{\Omega+H^{AB}\eta_{BA}} \ ,
\end{equation}
where $\Lambda$ and $\Omega$ are constants.
For (\ref{LH2ex}) we easily find
\begin{equation}
D_{AB}=
\frac{1}{2}\frac{\eta_{BA}}{\sqrt{\Omega+
H^{AB}\eta_{BA}}} \
\ , \quad
\tD^{AB}=2\eta^{AB}\sqrt{\Omega+
H^{AB}\eta_{BA}} \
\end{equation}
and hence
\begin{equation}
 J^{AB}\eta_{BA}=
\frac{V^{AB}\eta_{BA}-4\Omega K_{AB}\eta^{BA}}
{1+4 K_{AB}\eta^{BA}} \ .
\end{equation}
After some calculation we obtain
 the Hamiltonian
density for the scalar field in the form
\begin{equation}
\mH_T^\phi=\frac{\sqrt{\det h}}{16\pi G}
\sqrt{\Omega+V^{AB}\eta_{BA}}
\sqrt{1+4K_{AB}\eta^{BA}}-\frac{\sqrt{\det h}}{16\pi G}
\Lambda \ .
\end{equation}
Following the analysis
presented in previous section (\ref{fourth})
we find the gauge fixed Hamiltonian density
in the form
\begin{equation}
\mH_{fix}=\frac{\sqrt{\det h}}{16\pi G}
\sqrt{1+\left(\frac{16\pi G}{\sqrt{\det h}}\right)^2
\mH^{GR}_i\delta^{ij}\mH^{GR}_j
-\left(\frac{16\pi G}{\sqrt{\det h}}\mH^{GR}_T
-\Lambda\right)^2\frac{1}{\Omega+h^{ij}\delta_{ji}}}
 \
\end{equation}
and corresponding Lagrangian
\begin{eqnarray}\label{Lfix1}
L_{fix}&=&\int d^D\bx \left(\frac{1}{16\pi G}
\sqrt{-\det \hat{g}}(\hat{R}
+\Lambda)+BA+D^iC_i-\right.\nonumber \\
&-& \left. \frac{\sqrt{\det h}}{16\pi G}
\sqrt{1+\left(\frac{16\pi G}{\sqrt{\det h}}\right)^2
C_i\delta^{ij}C_j
-\left(\frac{16\pi G}{\sqrt{\det h}}A\right)^2\frac{1}{\Omega+h^{ij}\delta_{ji}}}
\right) \ . \nonumber \\
\end{eqnarray}

In this section we gave two explicit examples
of Hamiltonians and Lagrangians for massive gravity.
 Clearly it would be desirable to  understand properties of
these models further.

\section{Conclusion}\label{sixth}
This paper was devoted to the study of the Higgs mechanism
for gravity from the point of view of the Hamiltonian
formalism. We  performed the fixing of the
space-time diffeomorphism  and we argued that the
resulting Hamiltonian  corresponds to the Hamiltonian
  of the massive gravity. The my advantage
of our approach is that this  theory
is defined on the reduced phase space
spanned  by the  physical degrees of freedom
$h_{ij},p^{ij}$ only. On the other hand the price we pay for
this property is that it is difficult to find the
form of the gauge fixed Hamiltonian for general
potential $L(H^{AB})$. In fact,
we are not able to find explicit form of the gauge
fixed Hamiltonian for the specific form of the scalar
actions introduced in \cite{Kakushadze:2000zn,Kakushadze:2007dj,
Kakushadze:2007hf,Oda:2010gn,Chamseddine:2010ub,Oda:2010wn}.
On the other hand introducing additional auxiliary fields
we can determine  Lagrangian for massive gravity that has
the form of the ordinary General Relativity action with
specific potential terms that break diffeomorphism
invariance. Then we can ask the question
how this Lagrangian is related to the original Lagrangian
where we fix the gauge as in
\cite{Kakushadze:2000zn,Kakushadze:2007dj,
Kakushadze:2007hf,Oda:2010gn,Chamseddine:2010ub,Oda:2010wn}.
 One can hope that
these actions could be related by some fields redefinitions.
However finding this redefinition seems to be very complicated
due to the presence of the auxiliary fields $A,C_i$
 in the Lagrangian (\ref{Lfixfinal}) whose explicit
 integration out would lead to very obscure form of the
 Lagrangian.

The next important step in our investigation
would be to analyze  the spectrum of
fluctuations around the flat space-time background.
It would be also very interesting to study
the classical solutions corresponding to these
form of massive gravities.
 We hope to return to the analysis of
  these problems in future.

\vskip 5mm

 \noindent {\bf
Acknowledgements:}
I would like to thank CERN PH-TH for
generous hospitality and financial
 support during the course of this work.
 This work   was also
supported by the Czech Ministry of
Education under Contract No. MSM
0021622409. \vskip 5mm

\newpage

\end{document}